\documentclass[aps,amsfonts,nofootinbib]{revtex4}
\usepackage{epsfig}
\usepackage{graphicx}
\usepackage{amsmath}
\usepackage{amsbsy}
\usepackage{color}
\usepackage{epsfig}
\usepackage{graphicx}
\usepackage{amsmath}
\usepackage{amssymb}
\usepackage{bbm}
\usepackage{amsbsy}
\newcommand{\ee}{\end{equation}}
\newcommand{\bb}{\begin{equation}}
\newcommand{\eqb}{\begin{eqnarray}}
\newcommand{\eqf}{\end{eqnarray}}

\newcommand{\1}{{\'{\i}}}

\def\epp{e_{\scriptscriptstyle{(+)}}}
\def\emm{e_{\scriptscriptstyle{(-)}}}
\def\eppmm{e_{\scriptscriptstyle{(\pm)}}}
\def\chip{\chi_{\scriptscriptstyle{(+)}}}
\def\chim{\chi_{\scriptscriptstyle{(-)}}}
\def\chipm{\chi_{\scriptscriptstyle{(\pm)}}}

\def\1{\'{\i}}
\begin{document}
\title{Dark and Visible Photons as Source of CP Violation}
\author{J. Gamboa}
\email{jgamboa55@gmail.com}
\affiliation{Departamento de  F\'{\i}sica, Universidad de  Santiago de   
  Chile, Casilla 307, Santiago, Chile}
\author{F. Mendez}
\email{fernando.mendez@usach.cl}
\affiliation{Departamento de  F\'{\i}sica, Universidad de  Santiago de   
  Chile, Casilla 307, Santiago, Chile}

\begin{abstract}
The problem of excess gamma radiation in the center of galaxy is discussed assuming that the photon's production is dominated by two kinds of processes, the first one due to the conventional kinetic mixing term and, secondly, due to a  kinetic mixing  term violating the CP symmetry between dark and visible photons.  The CP violation symmetry between dark and visible sectors is not forbidden and, in principle, could be considered as an additional source of CP violation. The conversion probability between dark and visible photons is calculated and compared between both processes. The  processes violating CP are less significant but contribute  non-trivially to the excess gamma radiation.

\end{abstract}

\maketitle
The detection   of dark matter and  their constituents   are two  entangled  problems  subject to an intense theoretical and experimental scrutiny \cite{review1}. Generally  speaking, since we  still do  not  know all the symmetries which  fully describes  visible and dark matter, our predictive ability is still limited \cite{review2}. 

On the other hand, there are several observational results which could find a natural explanation from the particle physics point of view if both kind of matters and their interactions  find a place  in an unified model. One of them is the excess of gamma 
radiation from galaxies center. Most likely, as is discussed in many papers  \cite{many,adicional}, such photons are  produced by annihilation of pairs of 
dark matter ($d$) and anti-dark matter (${\bar d}$) in high density region producing gamma radiation \cite{planck,planck1}. 

However, although the hypothesis  $ {d}+{\bar{d}} \to {\mbox{photons}}$ is very reasonable, we cannot assure that in the dark sector there are processes 
that respecting all the obvious symmetries of standard particle physics. One of symmetries not necessarily fulfilled is CP symmetry (although CPT 
continue be an exact symmetry) and therefore, this is an issue that deserves to be studied.

Consider  dark ($X_\mu$) and visible photons ($A_\mu$) whose interactions may appear as kinetic mixing and can be implemented in different ways, one of them and  the best known,  is by adding to the $U(1)\times U'(1)$ Lagrangian,  the kinetic mixing term  \cite{holdom} (for recents developments see {\it e.g.} \cite{pospe})
\bb 
F_{\mu \nu} (A) F^{\mu \nu} (X)\equiv F(A)F(X). \label{idea1}
\ee  

 In order to explain this idea  let us assume a massive charged fermion ($\psi$), a visible photon ($A_\mu$) and a dark one ($X_\mu$) which interact by 
 means of  the following $U(1)\times U'(1)$ theory
 \eqb
 {\cal L} &=& {\bar \psi}( i {\partial \hspace{-.6em} \slash \hspace{.15em}} -
  {A \hspace{-.6em} \slash \hspace{.15em}}- m) \psi - 
 \frac{1}{4e^2}  F^2(A)  - \frac{1}{4} F^2 (X)  + 
 \frac{\chi}{2e} { F}(A)F (X) \nonumber 
 \\
 &=& {\cal L}_f + {\cal L} (A,X), \label{mod1}
\eqf
where $e$ is the charge of the fermion and 
\bb 
{\cal L} (A,X) = - 
 \frac{1}{4e^2}  F^2(A)  - \frac{1}{4} F^2 (X)  + 
 \frac{\chi}{2e} { F}(A)F (X). \label{mix}
 \ee

 The last Lagrangian can be diagonalized through the transformation $X'_\mu= X_\mu - (\chi/e )\, A_\mu$ and, after this transformation it turns out to be
 \bb 
 {\cal L} = -\frac{1}{4( {\emm})^2} F^(A) -\frac{1}{4} F^2 (X'), \label{mix2}
 \ee 
 and therefore (\ref{mod1}) becomes 
\begin{equation} 
{\cal L}= {\bar \psi}( i {\partial \hspace{-.6em} \slash \hspace{.15em}}  - {\emm}{A \hspace{-.6em} \slash \hspace{.15em}}- m) \psi 
 -\frac{1}{4} F^2(A)  - \frac{1}{4} F^2(X'), 
 \label{fermi}
\end{equation}
where $\chi$ is a real parameter and then, the effect of kinetic mixing is to redefine the electric charge 
\cite{redondo1}
\bb 
{ \emm}=\frac{e}{\sqrt{1-\chi^2}} 
\label{milli1}.
\ee
 
 Naively the redefinition of the electric charge  (\ref{milli1}) could be seen as a simple renormalization, however for $\chi ^ 2> 1$, the kinetic energy for $A$ changes sign,  ghosts appear and unitarity is lost. 
     
 In consequence, this rescaling (\ref{milli1}) is applicable only if $\chi^2 <1$. 
 If  $\chi=1$, $U(1)\times U'(1) \to U(1)$ and the action (\ref{fermi}) becomes equivalent to the standard QED.  
 
 The fact that not all the $\chi^2$  values are acceptable is uncomfortable and it seems reasonable to look for an alternative 
 that not only incorporate the kinetic mixing but also all possible values of  $\chi$. 
\vskip 0.20cm 

Then instead of (\ref{idea1}), we propose the kinetic mixing 
\bb 
F_{\mu \nu} (A) {\tilde F}^{\mu \nu} (X) \equiv F(A)\tilde{F}(X) =\tilde{F}(A)F(X), 
\label{idea2}
\ee
where the dual tensor is  ${\tilde F}^{\mu \nu} (A) = \frac{1}{2} \epsilon^{\mu \nu \rho \lambda} F_{\rho \lambda} (A)$.

The dynamics between both photons is given now by the Lagrangian 
 \bb 
{\cal L}_{\mbox{\tiny{gauge}}} = - \frac{1}{4e^2} F^2(A)  - \frac{1}{4} F^2 (X) + \frac{\chi}{2e } {\tilde F} (A)F (X).
\label{00}
\ee
Note first  that (\ref{idea2}) is gauge invariant but violates CP symmetry and therefore Lagrangian (\ref{00}) does. 
This last fact, however, is not a problem because there is no a physical basis for discarding this possibility. The coupling 
is also a boundary term which is irrelevant except for topologically nontrivial  field configurations of $X$ and/or  for a  space-time with   boundaries, 
and therefore we will keep it until the end of the calculation.


Under this assumptions, the inclusion of matter discussed in  (\ref{fermi}) is described now by  
\bb 
 {\cal L}= {\bar \psi}( i {\partial \hspace{-.6em} \slash \hspace{.15em}} - {A \hspace{-.6em} \slash \hspace{.15em}}- m) \psi -\frac{1}{4e^2}  F^{2} (A) - 
 \frac{1}{4}  F^{2} (X)) + \frac{\chi}{2e} F (A) {\tilde F} (X).
 \label{lag2}
\ee

It is possible to diagonalize (\ref{lag2}) and in doing so we only have to take care of the electromagnetic part ${\cal L}_{\mbox{\tiny{gauge}}}$ in 
(\ref{00}). In fact, let us perform a non-local transformation from the  gauge fields $\{A_\mu,X_\nu\}$ to $\{A_\mu,{A'}_\nu\}$ as follow
\begin{equation}
\label{nonlocal1}
F_{\mu\nu}({A'}) = F_{\mu\nu}(X) -\frac{\chi}{2e}\epsilon_{\mu\nu\lambda\rho}F^{\lambda\rho}(A),
\end{equation}
which satisfies 
\begin{equation}
\label{fp2}
F^2({A'})= F^{2}(X) -  \frac{\chi^2}{e^2}\,F^{2}(A) 
  - \frac{2\chi}{e}\,F(X){\tilde F}(A). 
\end{equation}
From this expression is direct to check that the Lagrangian (\ref{00}) is now decoupled and read
\begin{equation}
\label{01}
{\cal L}_{\mbox{\tiny{gauge}}}=-\frac{1}{4e^2}(1+\chi^2)F^2(A) -\frac{1}{4}F^2(A').
\end{equation}

In terms of local and gauge invariant quantities, namely the electric and magnetic fields, the Lagrangian (\ref{00}) is   
\bb 
{\cal L}_{\mbox{\tiny{gauge}}} = \frac{1}{2e^2} \left({\bf E}^2 -{\bf B}^2\right) +  \frac{1}{2} \left({\bf {E}}_X^2 -{\bf { B}}_X^2\right) -
\frac{\chi}{2e} \,\left( {\bf {E}}\cdot {\bf B}_X + {\bf { B}} \cdot {\bf E}_X \right).
 \label{cam}
\ee
where a subindex $X$ in the fields denote electromagnetic fields of the dark sector. That is  $E_i =F_{0i}(A)$,  $({E_X})_i = F_{0i}(X)$ and so on.

Then, from (\ref{nonlocal1}) we obtain the fields ${\bf E}',{\bf B}'$  
\eqb 
{ { \bf E}}' &=& {\bf E}_X - \frac{\chi}{e} \, {\bf {B}}, \nonumber 
\\ 
{{ \bf B}}' &=& {\bf B}_X + \frac{\chi}{e} \, {\bf { E}}, 
\label{diag2}
\eqf
and it is direct to show that the Lagrangian in (\ref{cam}) becomes 
\eqb
{\cal L}_{\mbox{\tiny{gauge}}} &=& \frac{1}{2e^2} (1+\chi^2) \left({\bf {E}}^2 -{\bf { B}}^2 \right) + \frac{1}{2} \left( { \bf E'}^2 - {\bf B'}^2 \right) \nonumber 
\\
&=& -\frac{1}{4{(\epp)}^2} F^2 (A) - \frac{1}{4} F^2(A'), 
\eqf
where the electric charge now is redefined as 
\bb 
{ e_{\scriptscriptstyle{(+)}}}=\frac{e}{\sqrt{1+\chi^2}}.  \label{milli12}
\ee

Note that this rescaled charge is valid for any value of $\chi$ and, therefore although it violates CP, the quantum theory is unitary. By the other hand,
Lagrangian (\ref{01}) is not CP invariant neither. This last statement can be verified directly through, for example, the relations in (\ref{diag2}), which 
are modified under CP transformation as follow
\begin{equation}
\begin{split}
{\bf E'}_{\mbox{\tiny{CP}}} &= -{\bf E'} -2\frac{\chi}{e}{\bf B},
\\
{\bf B'}_{\mbox{\tiny{CP}}} &=~ {\bf B'} -2\frac{\chi}{e}{\bf B}.
\end{split}
\end{equation}

Both  kinetic mixing procedures -- namely, couplings  (\ref{idea1}) and (\ref{idea2}) -- are described independently by one of the following  Lagrangians which includes fermionic matter
\bb 
{\cal L}_{\scriptscriptstyle{(\pm)}}= {\bar \psi}( i {\partial \hspace{-.6em} \slash \hspace{.15em}} -  {\eppmm}{A \hspace{-.6em} \slash \hspace{.15em}}- m) \psi -\frac{1}{4} F_{\mu \nu} (A) F^{\mu \nu} (A) - \frac{1}{4} F_{\mu \nu} (X') F^{\mu \nu} (X')), \label{ferm2}
\ee
with
\begin{equation}
\label{chargepm}
\eppmm =\frac{e}{\sqrt{1\pm({\chi_{\scriptscriptstyle{\pm}}})^2}},
\end{equation}
where we have chosen different couplings for the two different terms, namely $\chi_{\scriptscriptstyle{(+)}}$ for  (\ref{idea1}) and
 $\chi_{\scriptscriptstyle{(-)}}$ for (\ref{idea2}).

Even though this charge redefinition seems unobservable, one note that it happens  in the context of interaction of the dark sector with 
observable fields and  it  might  also  happen --because we do not know exactly the dark matter dynamics--  that the dark photon acquires 
mass ($m_{\gamma'}$),
{\it e.g.}  by spontaneous symmetry breaking, and in such conditions a photon oscillation
 process could take place \cite{okun,glashow}.

The conversion probability  for photon oscillation is therefore rescaled through the charge rescaling. This rescaling have an effect in the mass
of the hidden photon because $m_\gamma\propto \eppmm^2$ (depending on the coupling under discussion) and also in the probability itself which 
is proportional to $\eppmm^2$.
In fact, 
consider a beam of dark photons $\gamma'$ with energy $E_{\gamma'}$ and momentum ${\bf p}$ which oscillate to visible photons $\gamma$ with momentum 
${\bf p}$ and energy $E$ while traveling a time $T$. Then, the probability of conversion to visible photons, for the two couplings here considered, is \cite{okun}
\bb 
P^{(\pm)}_{\gamma' \to \gamma} \propto \frac{1}{1\pm (\chipm)^2} \sin^2 \left(\frac{|\Delta E| \,T}{2}\right), \label{osc12}
\ee 
where $|\Delta E| =|E -E_{\gamma'}|= ||{\bf p}| -\sqrt{{\bf p}^2 + m_{\gamma'}^2}| \approx \frac{m_{\gamma'}^2}{|{\bf p}|}= \frac{m_{\gamma'}^2}{E}$. 

The last formula implies that the ratio between both probabilities turn out to be 
\bb 
\frac{P^{(-)}_{\gamma' \to \gamma}}{P^{(+)}_{\gamma' \to \gamma}} =\frac{\sin^2\left(\frac{\kappa}{1-(\chim)^2}\right)}{1-(\chim)^2}
\frac{1+(\chip)^2}{\sin^2\left(\frac{\kappa}{1+(\chip)^2}\right)}, 
\ee
with $\kappa$ a constant.

Thus, we conclude that only for $|\chip| =|\chim|$ the conversion probabilities satisfy   $P^{(-)}_{\gamma' \to \gamma}> P^{(+)}_{\gamma' \to \gamma}$ and then the CP-violating
scenario is less favorable in the sense that this contribution might be less important to the excess of photons observed. However, for the 
general case  $\chip \neq \chim$, it is possible to have $P^{(-)}_{\gamma' \to \gamma} << P^{(+)}_{\gamma' \to \gamma}$  and then, the 
CP-violating conversion is the most relevant.

Thus, even in absence of precise observational data, we can conclude that excess of gamma radiation could be attributed to a combination of (\ref{idea1}) and CP violation processes (\ref{idea2}).

\medskip

\noindent\textbf{Acknowledgements}: We would like to thank Paola Arias for discussions. This  work was supported by grants from
FONDECYT-Chile grants 1130020 (J.G.) and 1140243 (F.M.).

\end{document}